%% file: celsym_v3.tex
\input myharvmac
\input amssym.def
\input amssym.tex
\noblackbox
\baselineskip=14.5pt

\def\comment#1{{}}

\def\zbar{{\bar z}}

\def\al{\alpha}

\def\ddz{\partial_z}\def\ddzb{\partial_{\bar z}}
\def\ddh{\partial_h}\def\ddhb{\partial_{\bar h}}

\newif\ifnref

\nreffalse

\input epsf

\def\figin{\epsfcheck\figin}\def\figins{\epsfcheck\figins}
\def\epsfcheck{\ifx\epsfbox\UnDeFiNeD
\message{(NO epsf.tex, FIGURES WILL BE IGNORED)}
\gdef\figin##1{\vskip2in}\gdef\figins##1{\hskip.5in}
\else\message{(FIGURES WILL BE INCLUDED)}%
\gdef\figin##1{##1}\gdef\figins##1{##1}\fi}
\def\DefWarn#1{}
\def\figinsert{\goodbreak\midinsert}  
\def\ifig#1#2#3{\DefWarn#1\xdef#1{Fig.~\the\figno}
\writedef{#1\leftbracket fig.\noexpand~\the\figno}%
\figinsert\figin{\centerline{#3}}\medskip\centerline{\vbox{\baselineskip12pt
\advance\hsize by -1truein\noindent\footnotefont\centerline{{\bf
Fig.~\the\figno}\ \sl #2}}}
\bigskip\endinsert\global\advance\figno by1}

\def\iifig#1#2#3#4{\DefWarn#1\xdef#1{Fig.~\the\figno}
\writedef{#1\leftbracket fig.\noexpand~\the\figno}%
\figinsert\figin{\centerline{#4}}\medskip\centerline{\vbox{\baselineskip12pt
\advance\hsize by -1truein\noindent\footnotefont\centerline{{\bf
Fig.~\the\figno}\ \ \sl #2}}}\smallskip\centerline{\vbox{\baselineskip12pt
\advance\hsize by -1truein\noindent\footnotefont\centerline{\ \ \ \sl #3}}}
\bigskip\endinsert\global\advance\figno by1}


\def\tilde{\widetilde}

\def\IR{ {\bf R}}






\lref\GreenXX{
  M.B.~Green and J.H.~Schwarz,
``Supersymmetrical Dual String Theory. 2. Vertices and Trees,''
Nucl.\ Phys.\ B {\bf 198}, 252 (1982).
}
\lref\GrossRR{
  D.J.~Gross, J.A.~Harvey, E.J.~Martinec and R.~Rohm,
``Heterotic String Theory. 2. The Interacting Heterotic String,''
Nucl.\ Phys.\ B {\bf 267}, 75 (1986).
}
\lref\LysovCSA{
  V.~Lysov, S.~Pasterski and A.~Strominger,
  ``Low’s Subleading Soft Theorem as a Symmetry of QED,''
Phys.\ Rev.\ Lett.\  {\bf 113}, no. 11, 111601 (2014).
[arXiv:1407.3814 [hep-th]].}
\lref\CheungIUB{
  C.~Cheung, A.~de la Fuente and R.~Sundrum,
  ``4D scattering amplitudes and asymptotic symmetries from 2D CFT,''
JHEP {\bf 1701}, 112 (2017).
[arXiv:1609.00732 [hep-th]].
}
\lref\HimwichDUG{
  E.~Himwich and A.~Strominger,
[arXiv:1901.01622 [hep-th]].
}
\lref\StiebergerWEA{
  S.~Stieberger,
``Closed superstring amplitudes, single-valued multiple zeta values and the Deligne associator,''
J.\ Phys.\ A {\bf 47}, 155401 (2014).
[arXiv:1310.3259 [hep-th]].
}

\lref\StiebergerLNG{
  S.~Stieberger and T.R.~Taylor,
  ``New relations for Einstein--Yang--Mills amplitudes,''
Nucl.\ Phys.\ B {\bf 913}, 151 (2016).
[arXiv:1606.09616 [hep-th]].
}

\lref\StromingerZOO{
  A.~Strominger,
  {\it Lectures on the Infrared Structure of Gravity and Gauge Theory},
 Princeton University Press (2018). [arXiv:1703.05448 [hep-th]].
}

\lref\PasterskiYLZ{
  S.~Pasterski, S.H.~Shao and A.~Strominger,
 ``Gluon Amplitudes as 2d Conformal Correlators,''
Phys.\ Rev.\ D {\bf 96}, no. 8, 085006 (2017).
[arXiv:1706.03917 [hep-th]].
}

\lref\LysovCSA{
  V.~Lysov, S.~Pasterski and A.~Strominger,
  ``Low's Subleading Soft Theorem as a Symmetry of QED,''
Phys.\ Rev.\ Lett.\  {\bf 113}, no. 11, 111601 (2014).
[arXiv:1407.3814 [hep-th]].
}

\lref\SchreiberJSR{
  A.~Schreiber, A.~Volovich and M.~Zlotnikov,
  ``Tree-level gluon amplitudes on the celestial sphere,''
Phys.\ Lett.\ B {\bf 781}, 349 (2018).
[arXiv:1711.08435 [hep-th]].
}
\lref\CheungIUB{
  C.~Cheung, A.~de la Fuente and R.~Sundrum,
  ``4D scattering amplitudes and asymptotic symmetries from 2D CFT,''
JHEP {\bf 1701}, 112 (2017).
[arXiv:1609.00732 [hep-th]].
}
\lref\HimwichDUG{
  E.~Himwich and A.~Strominger,
  ``Celestial Current Algebra from Low's Subleading Soft Theorem,''
[arXiv:1901.01622 [hep-th]].
}

\lref\PasterskiQVG{
  S.~Pasterski, S.H.~Shao and A.~Strominger,
  ``Flat Space Amplitudes and Conformal Symmetry of the Celestial Sphere,''
Phys.\ Rev.\ D {\bf 96}, no. 6, 065026 (2017).
[arXiv:1701.00049 [hep-th]].
}

\lref\TaylorSPH{
  T.R.~Taylor,
  ``A Course in Amplitudes,''
Phys.\ Rept.\  {\bf 691}, 1 (2017).
[arXiv:1703.05670 [hep-th]].
}

\lref\GrossAR{
  D.J.~Gross and P.F.~Mende,
``String Theory Beyond the Planck Scale,''
Nucl.\ Phys.\ B {\bf 303}, 407 (1988);
``The High-Energy Behavior of String Scattering Amplitudes,''
Phys.\ Lett.\ B {\bf 197}, 129 (1987).
}

\lref\fran{P. Di Francesco, P. Mathieu, D. S\'en\'echal, {\it Conformal Field Theory}, Springer-Verlag, New York (1997).}

\lref\PasterskiKQT{
  S.~Pasterski and S.H.~Shao,
  ``Conformal basis for flat space amplitudes,''
Phys.\ Rev.\ D {\bf 96}, no. 6, 065022 (2017).
[arXiv:1705.01027 [hep-th]].
}
\lref\FanUQY{
  W.~Fan, A.~Fotopoulos, S.~Stieberger and T.R.~Taylor,
``SV-map between Type I and Heterotic Sigma Models,''
Nucl.\ Phys.\ B {\bf 930}, 195 (2018).
[arXiv:1711.05821 [hep-th]].
}
\lref\StiebergerQJA{
  S.~Stieberger and T.R.~Taylor,
  ``Graviton Amplitudes from Collinear Limits of Gauge Amplitudes,''
Phys.\ Lett.\ B {\bf 744}, 160 (2015).
[arXiv:1502.00655 [hep-th]].
}
\lref\StiebergerCEA{
  S.~Stieberger and T.R.~Taylor,
  ``Graviton as a Pair of Collinear Gauge Bosons,''
Phys.\ Lett.\ B {\bf 739}, 457 (2014).
[arXiv:1409.4771 [hep-th]].
}

\lref\StiebergerVYA{
  S.~Stieberger and T.R.~Taylor,
``Disk Scattering of Open and Closed Strings (I),''
Nucl.\ Phys.\ B {\bf 903}, 104 (2016).
[arXiv:1510.01774 [hep-th]].
}
\lref\WittenNN{
  E.~Witten,
  ``Perturbative gauge theory as a string theory in twistor space,''
Commun.\ Math.\ Phys.\  {\bf 252}, 189 (2004).
[hep-th/0312171].
}
\lref\deBoerVF{
  J.~de Boer and S.N.~Solodukhin,
  ``A Holographic reduction of Minkowski space-time,''
Nucl.\ Phys.\ B {\bf 665}, 545 (2003).
[hep-th/0303006].
}
\lref\CheungIUB{
  C.~Cheung, A.~de la Fuente and R.~Sundrum,
  ``4D scattering amplitudes and asymptotic symmetries from 2D CFT,''
JHEP {\bf 1701}, 112 (2017).
[arXiv:1609.00732 [hep-th]].
}

\lref\StiebergerLNG{
  S.~Stieberger and T.R.~Taylor,
``New relations for Einstein-Yang-Mills amplitudes,''
Nucl.\ Phys.\ B {\bf 913}, 151 (2016).
[arXiv:1606.09616 [hep-th]].
}
\lref\KawaiXQ{
  H.~Kawai, D.C.~Lewellen and S.H.H.~Tye,
  ``A Relation Between Tree Amplitudes of Closed and Open Strings,''
Nucl.\ Phys.\ B {\bf 269}, 1 (1986).
}
\lref\SchlottererCXA{
  O.~Schlotterer,
``Amplitude relations in heterotic string theory and Einstein-Yang-Mills,''
JHEP {\bf 1611}, 074 (2016).
[arXiv:1608.00130 [hep-th]].
}
\lref\GRAV{
  S.~Stieberger,
 ``Constraints on Tree-Level Higher Order Gravitational Couplings in Superstring Theory,''
Phys.\ Rev.\ Lett.\  {\bf 106}, 111601 (2011)
[arXiv:0910.0180 [hep-th]].
}
\lref\StiebergerEDY{
  S.~Stieberger and T.R.~Taylor,
  ``Strings on Celestial Sphere,''
Nucl.\ Phys.\ B {\bf 935}, 388 (2018).
[arXiv:1806.05688 [hep-th]].
}
\lref\BernQJ{
  Z.~Bern, J.J.M.~Carrasco and H.~Johansson,
  ``New Relations for Gauge-Theory Amplitudes,''
Phys.\ Rev.\ D {\bf 78}, 085011 (2008).
[arXiv:0805.3993 [hep-ph]].
}

\Title{\vbox{\rightline{MPP--2018--292}
}}
{\vbox{\centerline{Symmetries of Celestial Amplitudes}
}}
\medskip
\centerline{Stephan Stieberger$^{a}$ ~,~ Tomasz R. Taylor$^{b}$}
\bigskip
\centerline{\it $^a$ Max--Planck--Institut f\"ur Physik}
\centerline{\it Werner--Heisenberg--Institut, 80805 M\"unchen, Germany}
\medskip
\centerline{\it  $^b$ Department of Physics}
\centerline{\it  Northeastern University, Boston, MA 02115, USA}
\vskip15pt

\vskip15pt

\medskip
\bigskip\bigskip\bigskip
\centerline{\bf Abstract}
\vskip .2in
\noindent
Celestial amplitudes provide holographic imprints of four-dimensional scattering processes in terms of conformal correlation functions on a two-dimensional sphere describing Minkowski space at null infinity. We construct the generators of Poincar\'e and conformal groups in the celestial representation and discuss how these symmetries are manifest in the amplitudes.
\noindent

\Date{}
\noindent
Celestial amplitudes provide holographic imprints of four-dimensional scattering processes in terms of conformal correlation functions on a two-dimensional celestial sphere describing Minkowski space at null infinity \StromingerZOO. In these amplitudes, Lorentz symmetry is realized as the $SL(2,C)$ conformal symmetry of the celestial sphere. They are particularly interesting in the soft limit, when one or more particles carry zero energy. In this limit, the well-known soft theorems can be interpreted as Ward identities of 2D CFT currents associated to asymptotic symmetries of four-dimensional spacetime \refs{\LysovCSA,\CheungIUB,\HimwichDUG}.   Beyond the soft limit, several examples of celestial amplitudes have been recently discussed in \refs{\PasterskiYLZ,\SchreiberJSR,\StiebergerEDY}.  The underlying 2D CFT may be very complicated but it is worth studying because it could lead to a holographic description of gauge theories and of (at least) some aspects of perturbative quantum gravity in asymptotically flat spacetimes. In the present work, we address the question how four-dimensional Poincar\'e and conformal symmetries are realized at the level of 2D celestial amplitudes. We construct the symmetry generators in the celestial representation. One of their interesting features is the presence of operators that shift
conformal dimensions. In particular, the momentum (spacetime translation) operators involve such shifts.

We first review the steps leading from ``old-fashioned'' to celestial amplitudes. We will be considering the scattering processes involving massless gauge bosons and gravitons.
Their asymptotic momenta, long before or after gauge/gravitational interactions take place, can be parametrized as
\eqn\mom{p_\mu=\omega q_\mu\ ,\quad {\rm with}~~ q_\mu={1\over 2}\ (1+|z|^2, z+\zbar,-i(z-\zbar),1-|z|^2)\ ,}
where $\omega$ are the (light-cone) energies and $(z,\zbar)$ are complex kinematic variables that determine momentum directions. On the other hand, the celestial sphere describing Minkowski spacetime at null infinity, from where these particles emerge and to where they head after they interact, is a Riemann sphere parameterized by complex coordinates. The starting point for constructing celestial amplitudes is the identification of kinematic variables $(z,\zbar)$ of Eq.\mom\ as the coordinates of points on celestial sphere. It follows that four-dimensional Lorentz group is realized as $SL(2,C)$ conformal symmetry of celestial sphere,
\eqn\conf{z\to{az+b\over cz+d}\qquad (ad-bc=1)\ .}
 In this framework, it is natural to replace the asymptotic {\bf in} and {\bf out} plane wave functions by the so-called conformal wave packets characterized by $(z,\zbar)$ and two-dimensional conformal weights ($h,\bar h$), with the conformal spin $J=h-\bar h$ identified as the helicity of the particle \PasterskiKQT. Their dimensions $\Delta= h+\bar h$  are restricted by the requirement of normalizability  to the so-called principal series with $\Delta=1+i\lambda$, $\lambda\in\IR$.

At the level of scattering amplitudes, the change of asymptotic {\bf in} and {\bf out} basis from plane waves to conformal packets is accomplished by Mellin transformations with respect to the energies:
\eqn\melt{\tilde{\cal A}_{\{h_n, \bar h_n\}}(z_n,\zbar_n)=\bigg(\prod_{n=1}^{N}
\int_0^\infty\omega_n^{\Delta_n-1}d\omega_n\bigg)\ \delta^{(4)}(\omega_1 q_1+\omega_2 q_2-\sum_{k=3}^N\omega_k q_k)\  {\cal M}(\omega_n,z_n,\zbar_n)\ ,}
with the dimensions $\Delta_n=h_n+\bar h_n$ dual to $\omega_n$.  Here,
${\cal M}$ are the $N$-particle invariant matrix elements describing particles 1 and 2 scattering into $N{-}2$ final particles.\foot{Our discussion applies though to all scattering channels. We omit the factor $i(2\pi)^4$. For notation, conventions and a general introduction into the subject, see \TaylorSPH. }  They depend on all quantum numbers, including internal gauge charges, and may contain some group-dependent (a.k.a.\ color) factors. In this case, we will be extracting purely kinematic ``partial'' (or ``stripped'') amplitudes associated to  individual Chan-Paton trace factors. Thus $\tilde{\cal A}(i,j,k,\dots)$ denotes the celestial amplitude associated to Tr$(T^{a_i}T^{a_j}T^{a_k}\cdots)$ \TaylorSPH.

The celestial amplitudes defined in Eq.\melt\
transform under conformal $SL(2,C)$ transformations like the correlation functions of $N$ conformal primary fields with weights $(h_n,\bar h_n)$. It is clear that Lorentz invariance of underlying amplitudes must be reflected in conformal Ward identities \fran. This helps in identifying the Lorentz generators
\eqn\lor{\eqalign{L_1\equiv M_{23}+iM_{10}& =(1-z^2)\ddz-2zh\ ,\qquad~
-M_{23}+iM_{10}=\bar L_1
\cr
L_2\equiv M_{20}+iM_{13}& =(1+z^2)\ddz+2zh\ ,\ \qquad
-M_{20}+iM_{13}=\bar L_2\cr
L_3\equiv M_{21}+iM_{03}& =2(z\ddz+h)\ ,\ \qquad\qquad~~
-M_{21}+iM_{03}=\bar L_3\ ,\cr}}
which obey the usual $\frak{su}(1,1)$ commutation relations
\eqn\lalg{\eqalign{
[L_1,L_2]&=2L_3\cr
[L_2,L_3]&=2L_1\cr
[L_3,L_1]&=-2L_2\ .\cr
}}
{}Indeed, as a consequence of conformal Ward identities, any $N$-point celestial amplitude satisfies the requirements of Lorentz invariance
\eqn\linv{{\cal L}_I\tilde{\cal A}_N=\bar {\cal L}_I\tilde{\cal A}_N=0\ ,\qquad {\cal L}_I=\sum_{n=1}^NL_{I,n}\ ,}
where $L_{I,n}\ ,~ I=1,2,3,$ are the Lorentz transformations \lor\ acting on the coordinates of $n$th particle.

In standard amplitudes, the momentum operator acts as multiplication by $p_\mu$ written in Eq.\mom. Multiplication by an $\omega$ energy factor yields a shift of conformal weights: $(h,\bar h)\to (h+1/2,\bar h+1/2)$, {\it cf}.\ Eq.\melt. Hence the momentum generators are realized as
\eqn\mome{\eqalign{P_0& =\big(1+|z|^2\big)e^{(\ddh+\ddhb)/2}
\cr
P_1&=(z+\zbar) e^{(\ddh+\ddhb)/2}\cr
P_2&=-i(z-\zbar)e^{(\ddh+\ddhb)/2}\cr
P_3& =\big(1-|z|^2\big)e^{(\ddh+\ddhb)/2}
}}
and the momentum conservation reads
\eqn\pinv{{\cal P}_\mu\tilde{\cal A}_N=0\ ,\qquad {\cal P}_\mu=P_{\mu,1}+P_{\mu,2}-\sum_{n=3}^NP_{\mu,n}\ ,}
where $P_{\mu,n}\ ,~ \mu=0,1,2,3,$ act on the coordinates of $n$th particle. The operators \lor\ and \mome\ generate the Poincar\'e group. It is easy to check that all celestial amplitudes written explicitly in Refs.\refs{\PasterskiYLZ,\SchreiberJSR,\StiebergerEDY} are Poincar\'e invariant; they satisfy Eqs.\linv\ and
\pinv. More details are given in the Appendix.

While shifting conformal weights may seem as a trivial operation, it affects the ultra-violet behaviour of Mellin transforms. Yang-Mills amplitudes are ``mariginally'' convergent, with the overall energy scale integral \refs{\PasterskiYLZ,\SchreiberJSR,\StiebergerEDY}
\eqn\deltl{\int_0^\infty \omega^{\big(\scriptstyle \sum_{n=1}^N\Delta_n-N-1\big)}d\omega=2\pi\delta \big(\textstyle\sum_{n=1}^N\lambda_n\big)}
(recall that $\Delta_n=1+i\lambda_n)$.
A shift of conformal dimension $\Delta_n\to\Delta_n+1$ induced by the momentum operator $P_{\mu,n}$ results in a linearly divergent integral.
As shown in Ref.\StiebergerEDY, such divergences can be avoided by treating the amplitudes as $\al'\to 0$ limits of superstring amplitudes. This works because superstring theory is ``supersoft'' in the ultra-violet: all scattering amplitudes are exponentially suppressed at high energies. In Ref.\StiebergerEDY, some four-point gravitational amplitudes have been discussed by using such a superstring embedding. Each power of the gravitational coupling constant (with mass dimension $-1$) brings an energy factor hence at the level of celestial amplitudes, it has the same effect as the momentum operator. Seen in this way,  celestial gravitational amplitudes appear as Yang-Mills amplitudes translated in space-time. As an example, the well-known relation~\StiebergerLNG\ between the Einstein-Yang-Mills amplitude with a single graviton $G$ and pure gauge amplitudes can be written as
\eqn\eymrel{g\,\tilde A(1,2,\dots, N, G^{\pm\pm})=\kappa\sum_{l=1}^{N-1}(\epsilon_G^{\pm\mu}{\cal X}_{\mu,l})\,\tilde A(1,2,\dots,l,G^{\pm},l+1,\dots,N),}
where
\eqn\eymrek{\qquad {\cal X}_{\mu,l}=P_{\mu,1}+P_{\mu,2}-\sum_{n=3}^lP_{\mu,n}\ ,}
while $g$ and $\kappa$ are the gauge and gravitational couplings, respectively. The polarization vectors are given by
\eqn\polar{\epsilon_{G\mu}^+={1\over\sqrt 2(z-w)}\ \big(1+\zbar w, w+\zbar,-i(w-\zbar),1-\zbar w\big)\ ,\qquad \epsilon_{G\mu}^{-}=(\epsilon_{G\mu}^{+ })^*\ ,}
where $w$ is a reference point on celestial sphere. The amplitude \eymrel\
does not depend on this point as a consequence of gauge invariance \StiebergerLNG\ which, in this case, follows from the Bern-Carrasco-Johansson relations \BernQJ. It is remarkable that the relations \eymrel\ hold in full-fledged heterotic superstring theory, to all orders in the $\al'$ expansion \SchlottererCXA.

At the tree-level, Yang-Mills theory is scale-invariant.
Accordingly, the tree-level helicity amplitudes are invariant
under four-dimensional conformal transformations \WittenNN. The celestial representation of special conformal generators can be deduced in a similar way as in Ref.\WittenNN. We find
\eqn\kas{\eqalign{K_0& =\big[\ddz\ddzb+(z\ddz+2h-1)(\zbar\ddzb+2\bar h-1)\big]e^{-(\ddh+\ddhb)/2}
\cr
K_1&=\big[(z\ddz+2h-1)\ddzb+(\zbar\ddzb+2\bar h-1)\ddz\big]e^{-(\ddh+\ddhb)/2}\cr
K_2&=-i\big[(z\ddz+2h-1)\ddzb-(\zbar\ddzb+2\bar h-1)\ddz\big]e^{-(\ddh+\ddhb)/2}\cr
K_3& =\big[\ddz\ddzb-(z\ddz+2h-1)(\zbar\ddzb+2\bar h-1)\big]
e^{-(\ddh+\ddhb)/2}
\ .}}
After computing all commutators one finds that indeed, the operators $L_I, ~\bar L_I, ~P_\mu, K_\mu$ of Eqs.{\lor, \mome\ and \kas}, supplemented by the dilatation generator
\eqn\dila{D=-i(h+\bar h-1)\ ,}
generate full conformal group \fran.
The simplest way of verifying Eq.\dila\ is by computing $[K_1, P_1]=2iD$.

Due to the complicated structure of the special conformal generators, {\it cf}.\ Eq.\kas, it is a tedious, although straightforward, exercise to show that the tree-level Yang-Mills amplitudes possess the symmetry \eqn\kasch{{\cal K}\tilde {\cal A}_{\rm YM}=0\ ,\qquad {\cal K}=K_{\mu,1}+K_{\mu,2}-\sum_{n=3}^NK_{\mu,l}\ .}
On the other hand, it is trivial to see that they are dilatation invariant.
Since $\Delta=1+i\lambda$,
\eqn\dilch{{\cal D}\tilde {\cal A}_{\rm YM}=0\ ,\qquad {\cal D}=\sum_{n=1}^N\lambda_n\ , }
due to the universal delta function \deltl\ present in all tree-level Yang-Mills amplitudes \refs{\PasterskiYLZ,\SchreiberJSR,\StiebergerEDY}.

In this work, we explained how Poincar\'e and conformal symmetries are realized in celestial amplitudes. In this formalism, gravitational amplitudes appear from space-time translations of pure gauge amplitudes, indicating that celestial CFT will be helpful in studying connections between gauge theories and gravity.\bigskip

\leftline{\noindent{\bf Acknowledgments}}\noindent
We are grateful to  Wei Fan, Angelos Fotopoulos, Sabrina Pasterski and Andy Strominger for useful conversations and communications.
This material is based in part upon work supported by the National Science Foundation
under Grant Number PHY--1620575.
Any opinions, findings, and conclusions or recommendations
expressed in this material are those of the authors and do not necessarily
reflect the views of the National Science Foundation.
\bigskip
\centerline{\noindent{\bf APPENDIX}}
We wish to expand the argument why celestial amplitudes are invariant under translations generated by the momentum operators \mome. To that end, we consider one particular generator,
\eqn\ppl{P_+={1\over 2}(P_0+
P_3) =e^{(\ddh+\ddhb)/2}\ ,\qquad {\cal P}_+=P_{+,1}+P_{+,2}-\sum_{n=3}^NP_{+,n}\ .}
The net effect of $P_+$ is to shift the conformal dimension $\Delta\to\Delta+1$ or equivalently $i\lambda\to i\lambda+1$.
Formally, ${\cal P}_+$ acting on the amplitude \melt, introduces the factor $(\omega_1+\omega_2-\sum_{k=3}^N\omega_k)$ under the Mellin integral. This factor is annihilated by the energy-conserving delta function, therefore
${\cal P}_+\tilde{\cal A}=0$. More caution should be exercised however because of possible convergence problems of Mellin transforms, therefore it is a good idea to have a closer look at some specific examples.\foot{We are grateful to Andy Strominger for suggesting this Appendix.} One such example is the tree-level, four-gluon Yang-Mills MHV amplitude that was Mellin-transformed into a celestial form in Refs.\refs{\PasterskiYLZ,\SchreiberJSR,\StiebergerEDY}. In the notation of Ref.\StiebergerEDY, it reads
\eqn\cgluf{\tilde{\cal A}(-,-,+,+)=A(z,\zbar,\lambda)J_0(\gamma)\ ,\qquad\gamma=i\sum_{n=1}^4\lambda_n\ , }
where
\eqn\cglug{A(z,\zbar,\lambda)= \delta(r-\bar r)\,\bigg({z_{24}\over \zbar_{13}}\bigg)^{i\lambda_1}\bigg({\zbar_{24}\over z_{13}}\bigg)^{i\lambda_3}\ \bigg({\zbar_{34}\over z_{12}}\bigg)^{i(\lambda_1+\lambda_2)}\bigg({z_{14}\over \zbar_{32}}\bigg)^{i(\lambda_2+\lambda_3)}{4r^3\over\zbar_{12}^2\, z_{34}^2}\ ,}
with $z_{ij}\equiv z_i-z_j$ and the real cross-ratio
\eqn\rat{r=\bar r ={z_{12}z_{34}\over z_{23}z_{41}}\ }
constrained to the kinematic domain of $r>1$. In Eq.\cgluf,
\eqn\jzero{J_0(\gamma)=\int_0^\infty \omega^{\gamma-1}d\omega\ ,}
see also Eq.\deltl. Acting on the amplitude \cgluf, each momentum generator $P_+$ shifts $\gamma\to\gamma+1$ and yields a factor rational in the $z$-coordinates. As a result,
\eqn\cgluh{{\cal P}_+\tilde{\cal A}(-,-,+,+)=\bigg[{z_{24}\zbar_{34}\over z_{12}\zbar_{13}}+{z_{14}\zbar_{34}\over z_{12}\zbar_{32}}-{z_{14}\zbar_{24}\over z_{13}\zbar_{32}}-1\bigg]A(z,\zbar,\lambda)J_0(\gamma+1)\ .}
After simple algebraic manipulations using the reality constraint $r=\bar r$, one finds that the expression inside the square bracket is zero. A similar argument can be repeated for the remaining momentum components, thus demonstrating translational invariance of the celestial MHV amplitude. Furthermore, it was shown in Ref.\StiebergerEDY\  that full-fledged superstring amplitudes describing four gluons in Type I and heterotic theories are given by expressions similar to \cgluf, with $J_0$ replaced by some more complicated functions of $\gamma$ and $r$. It is clear that the precise form of these functions does not affect Eq.\cgluh, therefore superstring amplitudes are also translationally invariant.
\listrefs
\end

%% file: myharvmac.tex
%
%
%
\def\unredoffs{} \def\redoffs{\voffset=-.31truein\hoffset=-.48truein}
\def\speclscape{}
%
%
%
%
%
\newbox\leftpage \newdimen\fullhsize \newdimen\hstitle \newdimen\hsbody
\tolerance=1000\hfuzz=2pt
\catcode`\@=11 
\ifx\hyperdef\UNd@FiNeD\def\hyperdef#1#2#3#4{#4}\def\hyperref#1#2#3#4{#4}\fi
\def\bigans{b }
\def\answ{b }
%
\ifx\answ\bigans\message{(This will come out unreduced.}
\magnification=1200\unredoffs\baselineskip=16pt plus 2pt minus 1pt
\hsbody=\hsize \hstitle=\hsize 
\else\message{(This will be reduced.} \let\l@r=L
\magnification=1000\baselineskip=16pt plus 2pt minus 1pt \vsize=7truein
\redoffs \hstitle=8truein\hsbody=4.75truein\fullhsize=10truein\hsize=\hsbody
\output={\ifnum\pageno=0 
  \shipout\vbox{\speclscape{\hsize\fullhsize\makeheadline}
    \hbox to \fullhsize{\hfill\pagebody\hfill}}\advancepageno
  \else
  \almostshipout{\leftline{\vbox{\pagebody\makefootline}}}\advancepageno
  \fi}
\def\almostshipout#1{\if L\l@r \count1=1 \message{[\the\count0.\the\count1]}
      \global\setbox\leftpage=#1 \global\let\l@r=R
 \else \count1=2
  \shipout\vbox{\speclscape{\hsize\fullhsize\makeheadline}
      \hbox to\fullhsize{\box\leftpage\hfil#1}}  \global\let\l@r=L\fi}
\fi
%
\newcount\yearltd\yearltd=\year\advance\yearltd by -2000

\def\Title#1#2{\nopagenumbers\abstractfont\hsize=\hstitle\rightline{#1}%
\vskip 1in\centerline{\titlefont #2}\abstractfont\vskip .5in\pageno=0}
\def\Date#1{\vfill\leftline{#1}\tenpoint\supereject\global\hsize=\hsbody%
\footline={\hss\tenrm\hyperdef\hypernoname{page}\folio\folio\hss}}%
%

\def\draftmode{\message{ DRAFTMODE }\def\draftdate{{\rm preliminary draft:
\number\month/\number\day/\number\yearltd\ \ \hourmin}}%
\headline={\hfil\draftdate}\writelabels\baselineskip=20pt plus 2pt minus 2pt
 {\count255=\time\divide\count255 by 60 \xdef\hourmin{\number\count255}
  \multiply\count255 by-60\advance\count255 by\time
  \xdef\hourmin{\hourmin:\ifnum\count255<10 0\fi\the\count255}}}
\def\nolabels{\def\wrlabeL##1{}\def\eqlabeL##1{}\def\reflabeL##1{}}
\def\writelabels{\def\wrlabeL##1{\leavevmode\vadjust{\rlap{\smash%
{\line{{\escapechar=` \hfill\rlap{\sevenrm\hskip.03in\string##1}}}}}}}%
\def\eqlabeL##1{{\escapechar-1\rlap{\sevenrm\hskip.05in\string##1}}}%
\def\reflabeL##1{\noexpand\llap{\noexpand\sevenrm\string\string\string##1}}}
\nolabels
%
\global\newcount\secno \global\secno=0
\global\newcount\meqno \global\meqno=1
\def\s@csym{}
\def\newsec#1{\global\advance\secno by1%
{\toks0{#1}\message{(\the\secno. \the\toks0)}}%
\global\subsecno=0\eqnres@t\let\s@csym\secsym\xdef\secn@m{\the\secno}\noindent
{\bf\hyperdef\hypernoname{section}{\the\secno}{\the\secno.} #1}%
\writetoca{{\string\hyperref{}{section}{\the\secno}{\it\the\secno.}} {{\it #1} }}%
\par\nobreak\medskip\nobreak}
\def\eqnres@t{\xdef\secsym{\the\secno.}\global\meqno=1\bigbreak\bigskip}
\def\sequentialequations{\def\eqnres@t{\bigbreak}}\xdef\secsym{}
\global\newcount\subsecno \global\subsecno=0
\def\subsec#1{\global\advance\subsecno by1%
{\toks0{#1}\message{(\s@csym\the\subsecno. \the\toks0)}}%
\ifnum\lastpenalty>9000\else\bigbreak\fi       \global\subsubsecno=0
\noindent{\it\hyperdef\hypernoname{subsection}{\secn@m.\the\subsecno}%
{\secn@m.\the\subsecno.} #1}\writetoca{\string\quad
{\string\hyperref{}{subsection}{\secn@m.\the\subsecno}{\secn@m.\the\subsecno.}}
{#1}}\par\nobreak\medskip\nobreak}
\def\appendix#1#2{\global\meqno=1\global\subsecno=0\xdef\secsym{\hbox{#1.}}%
\bigbreak\bigskip\noindent{\bf Appendix \hyperdef\hypernoname{appendix}{#1}%
{#1.} #2}{\toks0{(#1. #2)}\message{\the\toks0}}%
\xdef\s@csym{#1.}\xdef\secn@m{#1}%
\writetoca{\string\hyperref{}{appendix}{#1}{{\it Appendix} {\it #1.}} {\it #2}}%
\par\nobreak\medskip\nobreak}
%
%
\def\checkm@de#1#2{\ifmmode{\def\f@rst##1{##1}\hyperdef\hypernoname{equation}%
{#1}{#2}}\else\hyperref{}{equation}{#1}{#2}\fi}
\def\eqnn#1{\DefWarn#1\xdef #1{(\noexpand\relax\noexpand\checkm@de%
{\s@csym\the\meqno}{\secsym\the\meqno})}%
\wrlabeL#1\writedef{#1\leftbracket#1}\global\advance\meqno by1}
\def\f@rst#1{\c@t#1a\em@ark}\def\c@t#1#2\em@ark{#1}
\def\eqna#1{\DefWarn#1\wrlabeL{#1$\{\}$}%
\xdef #1##1{(\noexpand\relax\noexpand\checkm@de%
{\s@csym\the\meqno\noexpand\f@rst{##1}}{\hbox{$\secsym\the\meqno##1$}})}
\writedef{#1\numbersign1\leftbracket#1{\numbersign1}}\global\advance\meqno by1}
\def\eqn#1#2{\DefWarn#1%
\xdef #1{(\noexpand\hyperref{}{equation}{\s@csym\the\meqno}%
{\secsym\the\meqno})}$$#2\eqno(\hyperdef\hypernoname{equation}%
{\s@csym\the\meqno}{\secsym\the\meqno})\eqlabeL#1$$%
\writedef{#1\leftbracket#1}\global\advance\meqno by1}
\def\xeqn{\expandafter\xe@n}\def\xe@n(#1){#1}
\def\xeqna#1{\expandafter\xe@n#1}
\def\eqns#1{(\e@ns #1{\hbox{}})}
\def\e@ns#1{\ifx\UNd@FiNeD#1\message{eqnlabel \string#1 is undefined.}%
\xdef#1{(?.?)}\fi{\let\hyperref=\relax\xdef\next{#1}}%
\ifx\next\em@rk\def\next{}\else%
\ifx\next#1\xeqn#1\else\def\n@xt{#1}\ifx\n@xt\next#1\else\xeqna#1\fi
\fi\let\next=\e@ns\fi\next}

\def\DefWarn#1{\ifx\UNd@FiNeD#1\else
\immediate\write16{*** WARNING: the label \string#1 is already defined ***}\fi}
%
\newskip\footskip\footskip14pt plus 1pt minus 1pt 
\def\footnotefont{\ninepoint}\def\f@t#1{\footnotefont #1\@foot}
\def\f@@t{\baselineskip\footskip\bgroup\footnotefont\aftergroup\@foot\let\next}
\setbox\strutbox=\hbox{\vrule height9.5pt depth4.5pt width0pt}
\global\newcount\ftno \global\ftno=0
\def\foot{\global\advance\ftno by1\def\foot@rg{\hyperref{}{footnote}%
{\the\ftno}{\the\ftno}\xdef\foot@rg{\noexpand\hyperdef\noexpand\hypernoname%
{footnote}{\the\ftno}{\the\ftno}}}\footnote{$^{\foot@rg}$}}
%
\newwrite\ftfile
\def\footend{\def\foot{\global\advance\ftno by1\chardef\wfile=\ftfile
\hyperref{}{footnote}{\the\ftno}{$^{\the\ftno}$}%
\ifnum\ftno=1\immediate\openout\ftfile=\jobname.fts\fi%
\immediate\write\ftfile{\noexpand\smallskip%
\noexpand\item{\noexpand\hyperdef\noexpand\hypernoname{footnote}
{\the\ftno}{f\the\ftno}:\ }\pctsign}\findarg}%
\def\footatend{\vfill\eject\immediate\closeout\ftfile{\parindent=20pt
\centerline{\bf Footnotes}\nobreak\bigskip\input \jobname.fts }}}
\def\footatend{}
%
%
\global\newcount\refno \global\refno=1
\newwrite\rfile
\def\ref{[\hyperref{}{reference}{\the\refno}{\the\refno}]\nref}
\def\nref#1{\DefWarn#1%
\xdef#1{[\noexpand\hyperref{}{reference}{\the\refno}{\the\refno}]}%
\writedef{#1\leftbracket#1}%
\ifnum\refno=1\immediate\openout\rfile=\jobname.refs\fi
\chardef\wfile=\rfile\immediate\write\rfile{\noexpand\item{[\noexpand\hyperdef%
\noexpand\hypernoname{reference}{\the\refno}{\the\refno}]\ }%
\reflabeL{#1\hskip.31in}\pctsign}\global\advance\refno by1\findarg}
\def\findarg#1#{\begingroup\obeylines\newlinechar=`\^^M\pass@rg}
{\obeylines\gdef\pass@rg#1{\writ@line\relax #1^^M\hbox{}^^M}%
\gdef\writ@line#1^^M{\expandafter\toks0\expandafter{\striprel@x #1}%
\edef\next{\the\toks0}\ifx\next\em@rk\let\next=\endgroup\else\ifx\next\empty%
\else\immediate\write\wfile{\the\toks0}\fi\let\next=\writ@line\fi\next\relax}}
\def\striprel@x#1{} \def\em@rk{\hbox{}}
\def\lref{\begingroup\obeylines\lr@f}
\def\lr@f#1#2{\DefWarn#1\gdef#1{\let#1=\UNd@FiNeD\ref#1{#2}}\endgroup\unskip}

\def\addref#1{\immediate\write\rfile{\noexpand\item{}#1}} 
\def\listrefs{\footatend\vfill\supereject\immediate\closeout\rfile\writestoppt
\baselineskip=\footskip\centerline{{\bf References}}\bigskip{\parindent=20pt%
\frenchspacing\escapechar=` \input \jobname.refs\vfill\eject}\nonfrenchspacing}
\def\startrefs#1{\immediate\openout\rfile=\jobname.refs\refno=#1}
\def\xref{\expandafter\xr@f}\def\xr@f[#1]{#1}
\def\refs#1{\count255=1[\r@fs #1{\hbox{}}]}
\def\r@fs#1{\ifx\UNd@FiNeD#1\message{reflabel \string#1 is undefined.}%
\nref#1{need to supply reference \string#1.}\fi%
\vphantom{\hphantom{#1}}{\let\hyperref=\relax\xdef\next{#1}}%
\ifx\next\em@rk\def\next{}%
\else\ifx\next#1\ifodd\count255\relax\xref#1\count255=0\fi%
\else#1\count255=1\fi\let\next=\r@fs\fi\next}
%

%
\newwrite\ffile\global\newcount\figno \global\figno=1
\def\fig{fig.~\hyperref{}{figure}{\the\figno}{\the\figno}\nfig}
\def\nfig#1{\DefWarn#1%
\xdef#1{fig.~\noexpand\hyperref{}{figure}{\the\figno}{\the\figno}}%
\writedef{#1\leftbracket fig.\noexpand~\xfig#1}%
\ifnum\figno=1\immediate\openout\ffile=\jobname.figs\fi\chardef\wfile=\ffile%
{\let\hyperref=\relax
\immediate\write\ffile{\noexpand\medskip\noexpand\item{Fig.\ %
\noexpand\hyperdef\noexpand\hypernoname{figure}{\the\figno}{\the\figno}. }
\reflabeL{#1\hskip.55in}\pctsign}}\global\advance\figno by1\findarg}
\def\listfigs{\vfill\eject\immediate\closeout\ffile{\parindent40pt
\baselineskip14pt\centerline{{\bf Figure Captions}}\nobreak\medskip
\escapechar=` \input \jobname.figs\vfill\eject}}
\def\xfig{\expandafter\xf@g}\def\xf@g fig.\penalty\@M\ {}
\def\figs#1{figs.~\f@gs #1{\hbox{}}}
\def\f@gs#1{{\let\hyperref=\relax\xdef\next{#1}}\ifx\next\em@rk\def\next{}\else
\ifx\next#1\xfig #1\else#1\fi\let\next=\f@gs\fi\next}
\def\figin{\epsfcheck\figin}\def\figins{\epsfcheck\figins}
\def\epsfcheck{\ifx\epsfbox\UNd@FiNeD
\message{(NO epsf.tex, FIGURES WILL BE IGNORED)}
\gdef\figin##1{\vskip2in}\gdef\figins##1{\hskip.5in}
\else\message{(FIGURES WILL BE INCLUDED)}%
\gdef\figin##1{##1}\gdef\figins##1{##1}\fi}
\def\DefWarn#1{}
\def\figinsert{\goodbreak\midinsert}
\def\ifig#1#2#3{\DefWarn#1\xdef#1{Fig.~\noexpand\hyperref{}{figure}%
{\the\figno}{\the\figno}}\writedef{#1\leftbracket fig.\noexpand~\xfig#1}%
\figinsert\figin{\centerline{#3}}\medskip\centerline{\vbox{\baselineskip12pt
\advance\hsize by -1truein\noindent\wrlabeL{#1=#1}\footnotefont%
{\bf Fig.~\hyperdef\hypernoname{figure}{\the\figno}{\the\figno}:} #2}}
\bigskip\endinsert\global\advance\figno by1}
\newwrite\lfile
{\escapechar-1\xdef\pctsign{\string\%}\xdef\leftbracket{\string\{}
\xdef\rightbracket{\string\}}\xdef\numbersign{\string\#}}
\def\writedefs{\immediate\openout\lfile=\jobname.defs \def\writedef##1{%
{\let\hyperref=\relax\let\hyperdef=\relax\let\hypernoname=\relax
 \immediate\write\lfile{\string\def\string##1\rightbracket}}}}%
\def\writestop{\def\writestoppt{\immediate\write\lfile{\string\pageno
 \the\pageno\string\startrefs\leftbracket\the\refno\rightbracket
 \string\def\string\secsym\leftbracket\secsym\rightbracket
 \string\secno\the\secno\string\meqno\the\meqno}\immediate\closeout\lfile}}
\def\writestoppt{}\def\writedef#1{}
\def\seclab#1{\DefWarn#1%
\xdef #1{\noexpand\hyperref{}{section}{\the\secno}{\the\secno}}%
\writedef{#1\leftbracket#1}\wrlabeL{#1=#1}}
\def\subseclab#1{\DefWarn#1%
\xdef #1{\noexpand\hyperref{}{subsection}{\secn@m.\the\subsecno}%
{\secn@m.\the\subsecno}}\writedef{#1\leftbracket#1}\wrlabeL{#1=#1}}
\def\applab#1{\DefWarn#1%
\xdef #1{\noexpand\hyperref{}{appendix}{\secn@m}{\secn@m}}%
\writedef{#1\leftbracket#1}\wrlabeL{#1=#1}}
\newwrite\tfile \def\writetoca#1{}
\def\leaderfill{\leaders\hbox to 1em{\hss.\hss}\hfill}
\def\writetoc{\immediate\openout\tfile=\jobname.toc
   \def\writetoca##1{{\edef\next{\write\tfile{\noindent ##1
   \string\leaderfill {\string\hyperref{}{page}{\noexpand\number\pageno}%
                       {\noexpand\number\pageno}} \par}}\next}}}
\newread\ch@ckfile
\def\listtoc{\immediate\closeout\tfile\immediate\openin\ch@ckfile=\jobname.toc
\ifeof\ch@ckfile\message{no file \jobname.toc, no table of contents this pass}%
\else\closein\ch@ckfile\centerline{\bf Contents}\nobreak\medskip%
{\baselineskip=18.5pt  \footnotefont
\parskip=2pt\catcode`\@=12\input\jobname.toc
\catcode`\@=12\bigbreak\bigskip}\fi}
\catcode`\@=12 
%
\edef\tfontsize{\ifx\answ\bigans scaled\magstep3\else scaled\magstep4\fi}
\font\titlerm=cmr10 \tfontsize \font\titlerms=cmr7 \tfontsize
\font\titlermss=cmr5 \tfontsize \font\titlei=cmmi10 \tfontsize
\font\titleis=cmmi7 \tfontsize \font\titleiss=cmmi5 \tfontsize
\font\titlesy=cmsy10 \tfontsize \font\titlesys=cmsy7 \tfontsize
\font\titlesyss=cmsy5 \tfontsize \font\titleit=cmti10 \tfontsize
\skewchar\titlei='177 \skewchar\titleis='177 \skewchar\titleiss='177
\skewchar\titlesy='60 \skewchar\titlesys='60 \skewchar\titlesyss='60
\def\titlefont{\def\rm{\fam0\titlerm}
\textfont0=\titlerm \scriptfont0=\titlerms \scriptscriptfont0=\titlermss
\textfont1=\titlei \scriptfont1=\titleis \scriptscriptfont1=\titleiss
\textfont2=\titlesy \scriptfont2=\titlesys \scriptscriptfont2=\titlesyss
\textfont\itfam=\titleit \def\it{\fam\itfam\titleit}\rm}
 \ifx\answ\bigans\else scaled\magstep1\fi
\ifx\answ\bigans\def\abstractfont{\tenpoint}\else
\font\absit=cmti10 scaled \magstep1
\font\abssl=cmsl10 scaled \magstep1
\font\absrm=cmr10 scaled\magstep1 \font\absrms=cmr7 scaled\magstep1
\font\absrmss=cmr5 scaled\magstep1 \font\absi=cmmi10 scaled\magstep1
\font\absis=cmmi7 scaled\magstep1 \font\absiss=cmmi5 scaled\magstep1
\font\abssy=cmsy10 scaled\magstep1 \font\abssys=cmsy7 scaled\magstep1
\font\abssyss=cmsy5 scaled\magstep1 \font\absbf=cmbx10 scaled\magstep1
\skewchar\absi='177 \skewchar\absis='177 \skewchar\absiss='177
\skewchar\abssy='60 \skewchar\abssys='60 \skewchar\abssyss='60
\def\abstractfont{\def\rm{\fam0\absrm}
\textfont0=\absrm \scriptfont0=\absrms \scriptscriptfont0=\absrmss
\textfont1=\absi \scriptfont1=\absis \scriptscriptfont1=\absiss
\textfont2=\abssy \scriptfont2=\abssys \scriptscriptfont2=\abssyss
\textfont\itfam=\absit \def\it{\fam\itfam\absit}\def\footnotefont{\tenpoint}%
\textfont\slfam=\abssl \def\sl{\fam\slfam\abssl}%
\textfont\bffam=\absbf \def\bf{\fam\bffam\absbf}\rm}\fi
\def\tenpoint{\def\rm{\fam0\tenrm}
\textfont0=\tenrm \scriptfont0=\sevenrm \scriptscriptfont0=\fiverm
\textfont1=\teni  \scriptfont1=\seveni  \scriptscriptfont1=\fivei
\textfont2=\tensy \scriptfont2=\sevensy \scriptscriptfont2=\fivesy
\textfont\itfam=\tenit \def\it{\fam\itfam\tenit}\def\footnotefont{\ninepoint}%
\textfont\bffam=\tenbf \def\bf{\fam\bffam\tenbf}\def\sl{\fam\slfam\tensl}\rm}
\font\ninerm=cmr9 \font\sixrm=cmr6 \font\ninei=cmmi9 \font\sixi=cmmi6
\font\ninesy=cmsy9 \font\sixsy=cmsy6 \font\ninebf=cmbx9
\font\nineit=cmti9 \font\ninesl=cmsl9 \skewchar\ninei='177
\skewchar\sixi='177 \skewchar\ninesy='60 \skewchar\sixsy='60
\def\ninepoint{\def\rm{\fam0\ninerm}
\textfont0=\ninerm \scriptfont0=\sixrm \scriptscriptfont0=\fiverm
\textfont1=\ninei \scriptfont1=\sixi \scriptscriptfont1=\fivei
\textfont2=\ninesy \scriptfont2=\sixsy \scriptscriptfont2=\fivesy
\textfont\itfam=\ninei \def\it{\fam\itfam\nineit}\def\sl{\fam\slfam\ninesl}%
\textfont\bffam=\ninebf \def\bf{\fam\bffam\ninebf}\rm}
%
%
\def\noblackbox{\overfullrule=0pt}
\hyphenation{anom-aly anom-alies coun-ter-term coun-ter-terms}
\def\inv{^{\raise.15ex\hbox{${\scriptscriptstyle -}$}\kern-.05em 1}}

\def\Dsl{\,\raise.15ex\hbox{/}\mkern-13.5mu D} 
\def\dsl{\raise.15ex\hbox{/}\kern-.57em\partial}

\def\lspace{\ifx\answ\bigans{}\else\qquad\fi}
\def\lbspace{\ifx\answ\bigans{}\else\hskip-.2in\fi} 
\def\boxeqn#1{\vcenter{\vbox{\hrule\hbox{\vrule\kern3pt\vbox{\kern3pt
	\hbox{${\displaystyle #1}$}\kern3pt}\kern3pt\vrule}\hrule}}}
\def\mbox#1#2{\vcenter{\hrule \hbox{\vrule height#2in
		\kern#1in \vrule} \hrule}}  
%

\def\darr#1{\raise1.5ex\hbox{$\leftrightarrow$}\mkern-16.5mu #1}

\def\roughly#1{\raise.3ex\hbox{$#1$\kern-.75em\lower1ex\hbox{$\sim$}}}

\global\newcount\subsubsecno \global\subsubsecno=0
\def\subsubsec#1{\global\advance\subsubsecno by1%
{\toks0{#1}\message{(\the\secno\the\subsecno\the\subsubsecno. \the\toks0)}}%
\ifnum\lastpenalty>9000\else\bigbreak\fi
\noindent{\it\hyperdef\hypernoname{subsubsection}{\the\secno.\the\subsecno\the\subsubsecno}%
{\the\secno.\the\subsecno.\the\subsubsecno.} #1}
\par\nobreak\medskip\nobreak}
\def\boxit#1{\vbox{\hrule\hbox{\vrule\kern8pt
\vbox{\hbox{\kern8pt}\hbox{\vbox{#1}}\hbox{\kern8pt}}
\kern8pt\vrule}\hrule}}
\def\mathboxit#1{\vbox{\hrule\hbox{\vrule\kern8pt\vbox{\kern8pt
\hbox{$\displaystyle #1$}\kern8pt}\kern8pt\vrule}\hrule}}
\def\slashchar#1{\setbox0=\hbox{$#1$}           
   \dimen0=\wd0                                 
   \setbox1=\hbox{/} \dimen1=\wd1               
   \ifdim\dimen0>\dimen1                        
      \rlap{\hbox to \dimen0{\hfil/\hfil}}      
      #1                                        
   \else                                        
      \rlap{\hbox to \dimen1{\hfil$#1$\hfil}}   
      /                                         
   \fi}
\def\sqr#1#2{{\vcenter{\vbox{\hrule height.#2pt
         \hbox{\vrule width.#2pt height#1pt \kern#1pt
            \vrule width.#2pt}
         \hrule height.#2pt}}}}
